\documentclass[11pt,twoside]{report}
\usepackage{multicol,tabularx,array}
\pdfoutput=1
\usepackage[fit]{truncate}
\usepackage[T1]{fontenc}
\newcounter{dummy}

\newcommand{\BLKP}{
  \ifthenelse{\isodd{\value{page}}}{\relax}{\mbox{}\thispagestyle{empty}\newpage}}
\usepackage{geometry}
\usepackage{fancyhdr}
\newcommand{\ARTauthor}{~}
\newcommand{\ARTtitle}{~}
\usepackage{pdfpages}
\usepackage{hyperref}

\begin{document}
\pagestyle{empty}
\setlength{\fboxsep}{0pt}
\setlength{\fboxrule}{.0pt}
\thispagestyle{empty}
\setlength{\unitlength}{1mm}
\begin{picture}(0.001,0.001)
\put(135,8){CERN--2013--005}
\put(135,3){12 August 2013}
\put(0,-60){\includegraphics[width=15cm]{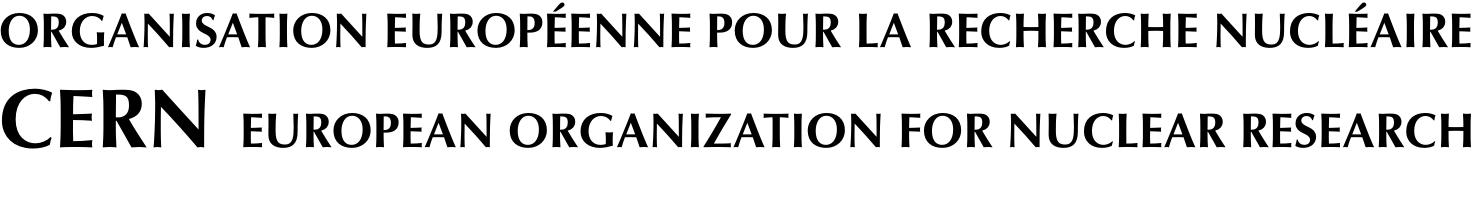}}
\put(15,-110){\LARGE\bfseries Fifty years of the CERN Proton Synchrotron}
\put(15,-125){\LARGE\bfseries Volume II}
\put(120,-185){\Large Editors:}
\put(138,-185){\Large S. Gilardoni}
\put(138,-191){\Large D. Manglunki}
\put(80,-250){\makebox(0,0){GENEVA}}
\put(80,-255){\makebox(0,0){2013}}
\end{picture}
\newpage
\thispagestyle{empty}
\mbox{}\\
\vfill
\begin{flushleft}
\begin{tabular}{@{}l@{~}l}
  ISBN & 978--92--9083--391--8 \\
  ISSN & 0007--8328\\ DOI  & 10.5170/CERN--2013--005\\ 
\end{tabular}\\[1mm]
Copyright \copyright{} CERN, 2013\\[1mm]
\raisebox{-1mm}{\includegraphics[height=12pt]{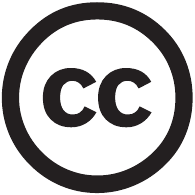}}
Creative Commons Attribution 3.0\\[1mm]
Knowledge transfer is an integral part of CERN's mission.\\[1mm]
CERN publishes this report Open Access under the Creative Commons
Attribution 3.0 license (\texttt{http://creativecommons.org/licenses/by/3.0/})
in order to permit its wide dissemination and use.\\[3mm]
This monograph should be cited as:\\[1mm]
Fifty years of the CERN Proton Synchrotron, volume II \\
edited by S. Gilardoni and D. Manglunki, CERN-2013-005 (CERN, Geneva, 2013), \\
DOI: 10.5170/CERN--2013--005\end{flushleft}

\pagenumbering{roman}

\pagestyle{plain}
\setcounter{page}{3}

\begin{center}
\mbox{}\\[3cm]
\bfseries\Large Dedication\\[1cm]
\end{center}
\noindent
The editors would like to express their gratitude to Dieter~M\"ohl, who passed away during the preparatory phase of this volume. This report is dedicated to him and to all the colleagues who, like him, contributed in the past with their cleverness, ingenuity, dedication and passion to the design and development of the CERN accelerators.\newpage\BLKP

\begin{center}
\mbox{}\\[15mm]
\bfseries\Large Abstract\\[1cm]
\end{center}
\noindent
This report sums up in two volumes the first 50 years of operation of
the CERN Proton Synchrotron. After an introduction on the genesis of
the machine, and a description of its magnet and powering systems, the 
first volume focuses on some of the many innovations in accelerator
physics and instrumentation that it has pioneered, such as transition
crossing, RF gymnastics, extractions, phase space tomography, or
transverse emittance measurement by wire scanners. The second volume 
describes the other machines in the PS complex: the proton linear
accelerators, the PS Booster, the LEP pre-injector, the heavy-ion
linac and accumulator, and the antiproton rings.
\newpage\BLKP

\begin{center}
\mbox{}\\[15mm]
\bfseries\Large Preface\\[1cm]
\end{center}
\addcontentsline{toc}{chapter}{Preface}
\refstepcounter{dummy}
\label{sec:preface}
\noindent
It was on 24 November 1959 that the proton beam in the CERN Proton
 Synchrotron was accelerated to a kinetic energy of 24 GeV. Thus the
 first strong-focusing proton synchrotron ever built has been
 faithfully serving the international physics community for 50
 years. It has been the subject of a virtually continuous upgrade
 boosting its intensity per pulse from \(10\sp{10}\) protons by more than three
 orders of magnitude to \(3\times10\sp{13}\) protons. Various injectors
 have been added and it has been modified such that, in addition to
 protons, light and heavy ions, positrons and electrons, as well as
 antiprotons could be accelerated or even decelerated often within the
 same supercycle. This would not have been possible had the initial
 design not been solid and sound allowing for maintainability,
 flexibility, and versatility and whose intrinsic potential was brought
 to fruition by the efforts and the ingenuity of generations of
 accelerator physicists, engineers, operators, and technicians. This report has been written to mark the fiftieth anniversary of the
 first operation of this unique accelerator. Volume I outlines the
 euphoric spirit in the European physics community in which such a bold
 design could be suggested, and gives an overview of the evolution of
 this unique accelerator described in a wealth of publications. This
 volume provides also a description in more depth of the outstanding
 achievements and highlights in its development. Volume II provides an
 overview of the injectors of the PS and of the accelerator system used
 for antiproton accumulation and storage, which has been closely
 associated with the PS.

\newpage\BLKP

\providecommand{\Titi}{Linac 1}
\providecommand{\Auti}{}
\providecommand{\Refi}{01-Linac1}

\providecommand{\Titii}{The Proton Synchrotron Booster (PSB)}
\providecommand{\Autii}{}
\providecommand{\Refii}{02-PSB}

\providecommand{\Titiii}{Linac 2}
\providecommand{\Autiii}{}
\providecommand{\Refiii}{03-Linac2}

\providecommand{\Titiv}{The Antiproton Accumulator, Collector and Decelerator Rings}
\providecommand{\Autiv}{}
\providecommand{\Refiv}{05-AA-AC-AD}

\providecommand{\Titv}{The Low-Energy Antiproton and Ion Rings LEAR and LEIR}
\providecommand{\Autv}{}
\providecommand{\Refv}{06-LEAR-LEIR}

\providecommand{\Titvi}{The LEP Pre-injector (LPI)}
\providecommand{\Autvi}{}
\providecommand{\Refvi}{04-LPI}

\providecommand{\Titvii}{Linac 3}
\providecommand{\Autvii}{}
\providecommand{\Refvii}{07-Linac3}

\begin{center}
\mbox{}\\[15mm]\bfseries\Large Contributors\\[25mm]
\end{center}
\refstepcounter{dummy}
\label{sec:contributors}
\begin{flushleft}
Jean-Paul~Burnet,
    Christian~Carli,
    Michel~Chanel,
    Roland~Garoby,
    Simone~Gilardoni,
    Massimo~Giovannozzi,
    Steven~Hancock,
    Helmut~Haseroth,
    Kurt~H\"ubner,
    Detlef~K\"uchler,
    Julian~Lewis,
    Alessandra~Lombardi,
    Django~Manglunki,
    Michel~Martini,
    Stephan~Maury,
    Elias~M\'etral,
    Dieter~M\"ohl,
    G\"unther~Plass,
    Louis~Rinolfi,
    Richard~Scrivens,
    Rende~Steerenberg,
    Charles~Steinbach,
    Maurizio~Vretenar,
    Thomas~Zickler
\end{flushleft}
%
%
%
\newpage\BLKP

\begin{center}
\mbox{}\\[15mm]\bfseries\Large Contents\\[25mm]
\end{center}
\begin{flushleft}
Preface 
   \dotfill~\pageref{sec:preface}\\[4mm]
List of contributors
   \dotfill~\pageref{sec:contributors}\\[4mm]
\Titi
   \dotfill~\pageref{S\Refi}\\[4mm]
\Titii
   \dotfill~\pageref{S\Refii}\\[4mm]
\Titiii
   \dotfill~\pageref{S\Refiii}\\[4mm]
\Titiv
   \dotfill~\pageref{S\Refiv}\\[4mm]
\Titv
   \dotfill~\pageref{S\Refv}\\[4mm]
\Titvi
   \dotfill~\pageref{S\Refvi}\\[4mm]
\Titvii
   \dotfill~\pageref{S\Refvii}\\[4mm]
\end{flushleft}
\newpage\BLKP


\cleardoublepage
\pagestyle{fancy}
\setcounter{page}{1}
\pagenumbering{arabic}
\newcommand{\Includeart}[3]{%
\renewcommand{\ARTauthor}{~}
\renewcommand{\ARTtitle}{~}
   \includepdf[frame,
               pages=1,
               noautoscale,
               rotateoversize=true,
               pagecommand={\pagestyle{fancy}},
               offset=0mm 0mm,
               addtotoc={1, section, 0, {\emph{#1.} #2},   S#3},
               trim=18mm 26mm 18mm 22mm, clip]
               {#3.pdf}
\renewcommand{\ARTauthor}{#2} 
\renewcommand{\ARTtitle}{#1}
   \includepdf[frame,
               pages=2-,
               rotateoversize=true,
               noautoscale,
               pagecommand={\pagestyle{fancy}},
               offset=0mm 0mm,
               trim=18mm 26mm 18mm 22mm, clip]
               {#3.pdf}
   \BLKP}

\Includeart{\Titi}{\Auti}{\Refi}
\Includeart{\Titii}{\Autii}{\Refii}
\Includeart{\Titiii}{\Autiii}{\Refiii}
\Includeart{\Titiv}{\Autiv}{\Refiv}
\Includeart{\Titv}{\Autv}{\Refv}
\Includeart{\Titvi}{\Autvi}{\Refvi}
\Includeart{\Titvii}{\Autvii}{\Refvii}

\cleardoublepage
\thispagestyle{empty}\ \clearpage
\thispagestyle{empty}\ \clearpage

\end{document}